\documentstyle[aps,amssymb,amsfonts,times,mathptm,12pt]{revtex}
\begin{document}
\small
\normalsize
\newcounter{saveeqn}
\newcommand{\alpheqn}{\setcounter{saveeqn}{\value{equation}}%
\setcounter{equation}{0}%
\renewcommand{\theequation}{\mbox{\arabic{saveeqn}-\alph{equation}}}}
\newcommand{\reseteqn}{\setcounter{equation}{\value{saveeqn}}%
\renewcommand{\theequation}{\arabic{equation}}}
\renewcommand{\thesection}{\Roman{section}}

\protect\newtheorem{principle}{Principle} [section]
\protect\newtheorem{theo}[principle]{Theorem}
\protect\newtheorem{prop}[principle]{Proposition}
\protect\newtheorem{lem}[principle]{Lemma}
\protect\newtheorem{co}[principle]{Corollary}
\protect\newtheorem{de}[principle]{Definition}
\newtheorem{ex}[principle]{Example}
\newtheorem{rema}[principle]{Remark}
\renewcommand{\baselinestretch}{1}
\small
\normalsize
\title{On unentangled Gleason theorems for quantum information theory}
\author{Oliver Rudolph \thanks{email: rudolph@starlab.net}}
\address{Physics Division, Starlab nv/sa, Boulevard Saint Michel
47, B-1040 Brussels, Belgium}
\setcounter{footnote}{0}
\author{J.D.~Maitland Wright \thanks{email: J.D.M.Wright@reading.ac.uk}}
\address{Analysis and Combinatorics Research Centre,
Mathematics,
University of Reading, 
Reading RG6 6AX, England}
\setcounter{footnote}{2}
\maketitle
\begin{abstract}
\noindent It is shown here that a strengthening of Wallach's
Unentangled Gleason Theorem can be obtained by applying results
of the present authors on generalised Gleason theorems for
quantum multi-measures
arising from investigations of quantum decoherence functionals.
\end{abstract}
\section{Introduction}
In an interesting recent paper Wallach \cite{Wallach}
obtained an \emph{unentangled} Gleason theorem.
His work was motivated by fundamental problems in
quantum information theory, in particular: to what extent do
local operations and measurements on multipartite quantum
systems suffice to guarantee the validity of a theorem of
Gleason-type and thus a Born-type rule for probabilities?
His positive result is formulated in terms of partially defined
frame functions, defined only on the \emph{unentangled states} of a
finite product of finite dimensional Hilbert spaces. We shall
show that Wallach's theorem, and also its generalisation
to infinite dimensions, can readily be derived from results which
were obtained by us in our investigations of generalised Gleason
theorems for quantum bi-measures and multi-measures
\cite{Wright95,RudolphW97,Wright98,RudolphW98,RudolphW99}.
The physical motivation for our earlier work arose from the
so-called histories approach to quantum mechanics
\cite{Isham94,IshamLS94}. Our more
general approach relies on the generalised Gleason theorem obtained by
Bunce and one of us \cite{BunceW92a,BunceW92b,BunceW92c}.
\section{Preliminaries}
Throughout this note ${\cal H}$ is a Hilbert space,
${\cal S}({\cal H})$ is the set of unit vectors in ${\cal H}$, and
the sets of projections,
compact operators or bounded operators are denoted by ${\cal P}({\cal H}),
{\cal K}({\cal H}),$
or ${\cal B}({\cal H})$
respectively. \\

A \emph{quantum measure} for ${\cal H}$ is a map
$m: {\cal P}({\cal H})
\to {\Bbb C}$ such that $m(p + q) =  m(p) + m(q)$ whenever
$p$ and $q$ are orthogonal.
If $m$ takes only positive values and $m(1) = 1$, then $m$ is
a \emph{quantum probability measure}. If, whenever
$\{p_i \}_{i \in I}$ is a family of mutually orthogonal projections,
$\sum_i m(p_i)$ is absolutely convergent and $m(\sum_i p_i) = \sum_i
m(p_i)$, then $m$ is said to be \emph{completely additive}. \\

The essential content of Gleason's original theorem \cite{Dvurecenskij93}
is that if $m$ is a positive, completely additive quantum measure on
${\cal P}({\cal H})$, then it has a unique extension to a positive
normal functional $\phi_m$ on ${\cal B}({\cal H})$, whenever the Hilbert
space ${\cal H}$ is not of dimension 2. It then follows from routine
functional analysis that there exists a unique positive, self-adjoint
trace class operator $T$ on ${\cal H}$ such that $\phi_m(x) =
{\mathrm{Tr}}(Tx)$ for each $x \in {\cal B}({\cal H})$, i.e.,
$m(p) = {\mathrm{Tr}}(Tp)$ for each $p \in {\cal P}({\cal H})$.
As a tool to help him prove his theorem, Gleason introduced the
notion of a frame function. A (positive) \emph{frame function} for
${\cal H}$ is a function $f: {\cal S}({\cal H}) \to {\mathbb{R}}^+$
such that there exists a real number $w$ (the \emph{weight} of $f$) such
that, for any orthonormal basis of ${\cal H}$, $\{ x_i \}_{i \in I}$,
$\sum_i f(x_i) = w$.  There is a bijective correspondence between
frame functions for $\mathcal{H}$
and positive, completely additive quantum
measures on ${\cal P}({\cal H})$, see \cite{Dvurecenskij93}.
\section{Unentangled frame functions and quantum multi-measures}
Let ${\cal H}_1, \cdots, {\cal H}_n$ be Hilbert spaces.
An \emph{unentangled} element of ${\cal H}_1 \otimes \cdots \otimes {\cal H}_n$
is a vector which can be expressed in the form $x_1 \otimes \cdots
\otimes x_n$. (Unentangled elements are sometimes referred to as
\emph{simple tensors}.) Let $\Sigma({\cal H}_1, \cdots, {\cal H}_n)$
be the set of all unentangled
vectors of norm 1 in ${\cal H}_1 \otimes \cdots \otimes {\cal H}_n$.
Every element in $\Sigma({\cal H}_1, \cdots, {\cal H}_n)$
can be expressed as a
tensor product of unit vectors in ${\cal H}_1, \cdots, {\cal H}_n$
respectively.
Following Wallach \cite{Wallach}, an
\emph{unentangled frame function} for ${\cal H}_1, \cdots, {\cal H}_n$
is a function $f: \Sigma({\cal H}_1, \cdots, {\cal H}_n) \to \mathbb{R}^+$
such that, for some positive real number $w$
(the \emph{weight} of $f$) whenever  $\{ \xi_i \}_{i \in I}$ is an
orthonormal basis of
${\cal H}_1 \otimes \cdots \otimes {\cal H}_n$ with each $\xi_i \in
\Sigma({\cal H}_1, \cdots, {\cal H}_n)$, then $\sum_i f(\xi_i) =
w$. The physical idea behind this definition is that the elements of
$\Sigma({\cal H}_1, \cdots, {\cal H}_n)$
represent the outcomes of elementary local operations or
measurements. \\

It turns out that unentangled frame functions have natural
links with quantum multi-measures. For the purposes of this note
we define a (positive) \emph{quantum multi-measure} for
${\cal H}_1, \cdots, {\cal H}_n$ to be a function \linebreak[3]
$m: {\cal P}({\cal H}_1)
\times \cdots \times {\cal P}({\cal H}_n) \to {\Bbb R}^+$, such that
$m$ is completely orthoadditive in each variable separately, see
\cite{RudolphW98}. (Our results in \cite{RudolphW98} apply to more
general, vector valued quantum multi-measures.)
When $n = 2$, a multi-measure is called a \emph{bi-measure}.
These arise naturally in the study of quantum decoherence functionals
\cite{IshamLS94,Wright95,RudolphW97,Wright98,RudolphW98,RudolphW99}.
\begin{lem} \label{l31}
Let ${\cal H}_1, \cdots, {\cal H}_n$ be Hilbert spaces, none of which
is of dimension 2. Let $m$ be a (positive) quantum multi-measure for
${\cal H}_1, \cdots, {\cal H}_n$. Then there exists a unique bounded,
multi-linear map $M : {\cal B}({\cal H}_1)
\times \cdots \times {\cal B}({\cal H}_n) \to {\Bbb C}$, such that
\[ M(p_1,p_2, \cdots, p_n) = m(p_1,p_2, \cdots, p_n) \mbox{ for
each } p_j \in {\cal B}({\cal H}_j). \]
Furthermore, given $r$, with $1 \leq r \leq n$ and assuming
$n \geq 2$, for each positive $x_j \in {\cal B}({\cal H}_j)$,
with $1 \leq j \leq n$ and $j \neq r$, the map
$y \mapsto M(x_1, \cdots, x_{r-1},y,x_{r+1}, \cdots, x_n)$ is a positive
normal functional on ${\cal B}({\cal H}_r)$.
\end{lem}
\emph{Proof}: The existence and uniqueness of $M$ is a consequence of
results obtained in \cite{RudolphW98}.
Whenever $x$ is a positive operator in ${\mathcal{B}}({\mathcal{H}})$,
there exists a sequence of commuting projections $\{ p_j \}_{j = 1, 2,
\cdots}$ such that $x = \Vert x \Vert \sum_j \frac{1}{2^j} p_j$ (for
a proof see, e.g., \cite{Pedersen79} page 27).
This observation, together with the positivity of $m$, shows that if
$x_r$ is positive for $r = 1,2, \cdots, n$, then
$M(x_1, \cdots,x_n) \geq 0$. It now follows from the results of
\cite{RudolphW98} that given $r$, with $1 \leq r \leq n$ and assuming
$n \geq 2$, for each positive $x_j \in {\mathcal{B}}({\mathcal{H}}_j)$,
with $1 \leq j \leq n$ and $j \neq r$, the map
$y \mapsto M(x_1, \cdots, x_{r-1},y,x_{r+1}, \cdots, x_n)$
is a positive normal functional on
${\mathcal{B}}({\mathcal{H}}_r)$. $\Box$ \\

Let us recall that the algebraic tensor product
${\cal B}({\cal H}_1) \otimes_{\rm alg} \cdots \otimes_{\rm alg}
{\cal B}({\cal H}_n)$ may be identified with the linear span of
$\{ x_1 \otimes x_2 \otimes \cdots \otimes x_n: x_j \in
{\cal B}({\cal H}_j) \}$ in (the von Neumann tensor product)
${\cal B}({\cal H}_1
\otimes \cdots \otimes {\cal H}_n) =
{\cal B}({\cal H}_1) \otimes \cdots \otimes {\mathcal{B}}({\cal H}_n)$.
Let $m$ and $M$ be as in Lemma \ref{l31},
then by the basic property of the algebraic tensor product,
there exists a unique linear functional ${\mathfrak{M}}$ on
${\cal B}({\cal H}_1)
\otimes_{\rm alg} \cdots \otimes_{\rm alg} {\cal B}({\cal H}_n)$
such that \\ 
${\mathfrak{M}}(x_1 \otimes x_2 \otimes \cdots \otimes x_n) =
M(x_1,x_2, \cdots, x_n)$.
\begin{co} Let ${\cal H}_1, \cdots, {\cal H}_n$ be finite dimensional
Hilbert spaces,
none of which has dimension 2. Let $m$ be a positive quantum multi-measure
for ${\cal H}_1, \cdots, {\cal H}_n$. Then there exists an unentangled
frame function $f$ for ${\cal H}_1, \cdots, {\cal H}_n$ such that, whenever
$\nu_1 \otimes \nu_2 \otimes \cdots \otimes \nu_n$ is in
$\Sigma({\cal H}_1, \cdots, {\cal H}_n)$ and $p_j$ is the projection of
${\cal H}_j$ onto the one-dimensional subspace generated by $\nu_j$,
\[ f(\nu_1 \otimes \nu_2 \otimes \cdots \otimes \nu_n) =
m(p_1,p_2, \cdots, p_n). \] \end{co}
\emph{Proof}: Fix  a unit vector $\nu_j$ in ${\cal H}_j$ for
$j = 1,2, \cdots, n$. Then the projection from ${\cal H}_1
\otimes \cdots \otimes {\cal H}_n$ onto the subspace spanned by
$\nu_1 \otimes \nu_2 \otimes \cdots \otimes \nu_n$ can be identified
with the projection $p_1 \otimes p_2 \otimes \cdots \otimes p_n$ in
${\cal B}({\cal H}_1 \otimes \cdots \otimes
{\cal H}_n) = {\cal B}({\cal H}_1) \otimes \cdots \otimes
{\cal B}({\cal H}_n)$. Define $f(\nu_1 \otimes \nu_2
\otimes \cdots \otimes \nu_n)$ to be
${\mathfrak{M}}(p_1 \otimes p_2 \otimes \cdots \otimes p_n) =
M(p_1,p_2, \cdots,p_n) = m(p_1,p_2, \cdots,p_n)$. $\Box$ \\

The following technical lemma allows us to associate a canonical
multi-measure with each unentangled frame function.
\begin{lem} \label{l33}
Let ${\cal H}_1, \cdots, {\cal H}_n$ be Hilbert spaces
of arbitrary dimension and
let $f: \Sigma({\cal H}_1, \cdots, {\cal H}_n) \to \mathbb{R}^+$
be an unentangled frame function. Then there is a (positive,
completely additive) quantum multi-measure $m$ for ${\cal H}_1,
\cdots, {\cal H}_n$ such that whenever
$\nu_1 \otimes \nu_2 \otimes \cdots \otimes \nu_n$ is in
$\Sigma({\cal H}_1, \cdots, {\cal H}_n)$ and $p_j$ is the projection of
${\cal H}_j$ onto the one-dimensional subspace generated by $\nu_j$,
\[ m(p_1,p_2, \cdots, p_n) =
f(\nu_1 \otimes \nu_2 \otimes \cdots \otimes \nu_n). \] \end{lem}
\emph{Proof}: To simplify our notation we shall prove this for
$n = 2$, but the method is perfectly general.

Let $e_1$ and $e_2$ be projections in ${\mathcal{P}}({\mathcal{H}}_1)$
and ${\mathcal{P}}({\mathcal{H}}_2)$, respectively. Let $E_1$ and $E_2$
be the subspaces of ${\mathcal{H}}_1$ and ${\mathcal{H}}_2$ which are
the respective ranges of $e_1$ and $e_2$. Let $\{ \xi_j \}_{j \in J}$
and $\{ \psi_i \}_{i \in I}$ be orthonormal bases of $E_1$ and $E_2$,
respectively. We wish to define $m(e_1,e_2)$ to be \[
\sum_{j \in J} \sum_{i \in I} f(\xi_j \otimes \psi_i). \]
The only difficulty here is that we do not know that this number is
independent of the choice of orthonormal bases for $E_1$ and $E_2$,
respectively. To establish this we argue as follows.

Let $w$ be the weight of $f$. Let $\{ \xi_j \}_{j \in J^\perp}$ and
$\{ \psi_i \}_{i \in I^\perp}$ be orthonormal bases for $E_1^\perp$ and
$E_2^\perp$, respectively. Then $\{ \xi_j \otimes \psi_i \}_{j \in J \cup
J^\perp, i \in I \cup I^\perp}$ is an orthonormal basis for
${\mathcal{H}}_1 \otimes {\mathcal{H}}_2$. So
\[ \sum_{(j,i) \in J \times I} f(\xi_j \otimes \psi_i) + \sum_{(j,i) \in
J^\perp \times (I \cup I^\perp)} f(\xi_j \otimes \psi_i) + \sum_{(j,i) \in
(J \cup J^\perp) \times I^\perp} f(\xi_j \otimes \psi_i) = w. \]
Let $\{ \xi_j' \}_{j \in J}$
and $\{ \psi_i' \}_{i \in I}$  be orthonormal bases of $E_1$ and $E_2$,
respectively. Then
\[ \sum_{(j,i) \in J \times I} f(\xi_j' \otimes \psi_i') + \sum_{(j,i) \in
J^\perp \times (I \cup I^\perp)} f(\xi_j \otimes \psi_i) + \sum_{(j,i) \in
(J \cup J^\perp) \times I^\perp} f(\xi_j \otimes \psi_i) = w. \]
Hence \begin{eqnarray*} \sum_{(j,i) \in J \times I}
f(\xi_j' \otimes \psi_i') & = &
w - \sum_{(j,i) \in
J^\perp \times (I \cup I^\perp)} f(\xi_j \otimes \psi_i) - \sum_{(j,i) \in
(J \cup J^\perp) \times I^\perp} f(\xi_j \otimes \psi_i) \\
& = & \sum_{(j,i) \in J \times I}
f(\xi_j \otimes \psi_i). \end{eqnarray*}
So $m$ is well-defined. It is straightforward to verify that $m$ has
all the required properties. $\Box$ \\

\noindent\emph{Remark}: In the above argument we made essential use of the
property that $f$ is an unentangled frame function. Suppose that
we only knew that $f$ satisfied the weaker property: for some
positive real number $w$ whenever $\{ \xi_i \}_{i \in I}$ is a product
orthonormal basis of ${\cal H}_1 \otimes \cdots \otimes
{\cal H}_n$, then $\sum_i f(\xi_i) = w$. Then the proof of
the preceding lemma would break down. This throws fresh light on the
counterexample constructed in Proposition 5 in \cite{Wallach}. \\

In our investigations on quantum decoherence functionals we were
led to obtain results on generalised quantum bi-measures and
multi-measures \cite{Wright95,RudolphW97,Wright98,RudolphW98,RudolphW99}.
The statement of the next theorem is Wallach's Theorem 1 \cite{Wallach}.
Our proof shows that Wallach's Theorem  is a natural consequence of
our earlier results on quantum multi-measures.

\begin{prop}[Wallach, Theorem 1 \cite{Wallach}] \label{w}
Let ${\cal H}_1, \cdots, {\cal H}_n$ be finite dimensional Hilbert
spaces, each of dimension
at least 3. Let $f: \Sigma({\cal H}_1, \cdots, {\cal H}_n) \to
\mathbb{R}^+$ be an unentangled frame function. Then there exists a
self-adjoint operator $T$ in ${\cal B}({\cal H}_1 \otimes \cdots \otimes
{\cal H}_n)$ such that whenever $\nu_1 \otimes \nu_2 \otimes \cdots
\otimes \nu_n$ is in $\Sigma({\cal H}_1, \cdots, {\cal H}_n)$ and
$p_j$ is the projection of ${\cal H}_j$ onto the one-dimensional
subspace generated by $\nu_j$, \[
f(\nu_1 \otimes \nu_2 \otimes \cdots
\otimes \nu_n) = {\mathrm{Tr}}((p_1 \otimes p_2 \otimes \cdots \otimes
p_n)T). \] \end{prop}
\emph{Proof}: Since each of the Hilbert spaces ${\cal H}_1,
\cdots, {\cal H}_n$ is finite dimensional,
${\cal B}({\cal H}_1) \otimes \cdots \otimes {\cal B}(
{\cal H}_n) = {\cal B}({\cal H}_1) \otimes_{\mathrm{alg}} \cdots
\otimes_{\mathrm{alg}} {\cal B}({\cal H}_n)$. Let $m$ be the quantum
multi-measure constructed from $f$ as in Lemma \ref{l33}.
Let $\mathfrak{M}$ be the linear functional on
${\cal B}({\cal H}_1) \otimes_{\mathrm{alg}} \cdots
\otimes_{\mathrm{alg}} {\cal B}({\cal H}_n)
= {\cal B}({\cal H}_1) \otimes \cdots
\otimes {\cal B}({\cal H}_n) = {\cal B}({\cal H}_1 \otimes \cdots
\otimes {\cal H}_n)$ such that
${\mathfrak{M}}(q_1 \otimes q_2 \otimes \cdots \otimes q_n) =
M(q_1,q_2, \cdots, q_n) = m(q_1,q_2, \cdots, q_n)$
for each $q_j \in {\cal P}({\cal H}_j)$.
Since $\mathfrak{M}$ is a linear functional on a finite dimensional
space, it is bounded. Hence there is a unique bounded operator $T$
in ${\cal B}({\cal H}_1 \otimes \cdots
\otimes {\cal H}_n)$ such that ${\mathfrak{M}}(x) = {\mathrm{Tr}}(xT)$
for all $x$. Thus
\begin{equation} \label{beepbeep} f(\nu_1 \otimes \nu_2 \otimes \cdots
\otimes \nu_n) = {\mathfrak{M}}(p_1 \otimes p_2 \otimes \cdots \otimes p_n)
= {\mathrm{Tr}}((p_1 \otimes p_2 \otimes \cdots \otimes p_n)T).
\end{equation}
On taking complex conjugates of the Equation (\ref{beepbeep}) we find
that $T$
may be replaced by $T^*$. So in (\ref{beepbeep}) we may replace $T$ by
$\frac{1}{2}
(T+T^*)$. Hence we may suppose in (\ref{beepbeep}) that $T$ is
self-adjoint. $\Box$ \\

The work of \cite{RudolphW98,RudolphW99} shows that Wallach's Theorem
can be generalised to the situation where the Hilbert spaces are not
required to be finite dimensional provided an appropriate boundedness
condition is imposed. More precisely:
\begin{theo} Let ${\cal H}_1, \cdots, {\cal H}_n$ be Hilbert spaces,
each of dimension at least 3. \linebreak[3] \\
Let $f: \Sigma({\cal H}_1, \cdots, {\cal H}_n) \to
\mathbb{R}^+$ be an unentangled frame function. Let ${\mathfrak{M}}$ be
the associated linear functional on
${\cal B}({\cal H}_1) \otimes_{\mathrm{alg}} \cdots
\otimes_{\mathrm{alg}} {\cal B}({\cal H}_n)$.
If the restriction of $\mathfrak{M}$ to ${\cal K}({\cal H}_1)
\otimes_{\mathrm{alg}} \cdots
\otimes_{\mathrm{alg}} {\mathcal{K}}({\cal H}_n)$ is bounded, then
 there exists a unique bounded  self-adjoint, trace class operator $T$ in
  ${\cal B}({\cal H}_1 \otimes \cdots
\otimes {\cal H}_n)$ such that whenever $\nu_1 \otimes \nu_2 \otimes \cdots
\otimes \nu_n$ is in
  $\Sigma({\cal H}_1, \cdots, {\cal H}_n)$ and $p_j$ is the projection
  of ${\cal H}_j$ onto the
  one-dimensional subspace generated by $\nu_j$,
 \[ f(\nu_1 \otimes \nu_2 \otimes \cdots
\otimes \nu_n) = {\mathrm{Tr}}((p_1 \otimes p_2 \otimes \cdots \otimes
p_n)T). \] \label{t35} \end{theo}
\emph{Proof}:
Let ${\mathfrak{M}}_0$ be the restriction of ${\mathfrak{M}}$ to
\linebreak[3] \mbox{${\cal K}({\cal H}_1)
\otimes_{{\mathrm{alg}}} \cdots \otimes_{{\mathrm{alg}}}
{\mathcal{K}}({\cal H}_n)$}. By hypothesis ${\mathfrak{M}}_0$ is bounded
and so has a unique bounded extension, also denoted by ${\mathfrak{M}}_0$,
to \linebreak[3] \mbox{${\cal K}({\cal H}_1
\otimes \cdots \otimes {\cal H}_n)$}.
By standard functional analysis, there exists a trace class operator
$T$ such that ${\mathfrak{M}}_0(z) = {\mathrm{Tr}}(zT)$ for each $z$ in
\linebreak[3] \mbox{${\cal K}({\cal H}_1
\otimes \cdots \otimes {\cal H}_n)$}. Since each one-dimensional
projection in ${\mathcal{B}}({\mathcal{H}}_j)$ is in
${\mathcal{K}}({\mathcal{H}}_j)$, \[
m(p_1,p_2, \cdots, p_n) = M(p_1,p_2, \cdots,p_n) =
{\mathfrak{M}}_0(p_1 \otimes p_2 \otimes \cdots \otimes p_n) =
{\mathrm{Tr}}((p_1 \otimes p_2 \otimes \cdots \otimes p_n)T). \]
So \[ f(\nu_1 \otimes \nu_2 \otimes \cdots \otimes \nu_n) =
{\mathrm{Tr}}((p_1 \otimes p_2 \otimes \cdots \otimes p_n)T) =
\langle T (\nu_1 \otimes \nu_2 \otimes \cdots \otimes \nu_n),
\nu_1 \otimes \nu_2 \otimes \cdots \otimes \nu_n \rangle, \]
where $\langle \cdot, \cdot \rangle$ denotes the inner product on
${\cal H}_1
\otimes \cdots \otimes {\cal H}_n.$
It now follows from Lemma 5.9 [6] that $T$ is unique. But, arguing
as in the proof of Proposition \ref{w}, we can replace
$T$ by $\frac{1}{2}(T+T^*)$. So, by uniqueness, $T$ is
self-adjoint. $\Box$ \\

\noindent\emph{Remark}: When the spaces ${\cal H}_1, \cdots, {\cal H}_n$ are
finite dimensional, then the boundedness condition of Theorem \ref{t35}
is automatically satisfied. So Wallach's Theorem is a corollary of
Theorem \ref{t35} which, in turn, follows from the work of
\cite{RudolphW98,RudolphW99}.

Moreover, it can be shown along the lines of \cite{RudolphW97} that,
for $n = 2$, there exists a self-adjoint operator $T$, not necessarily
of trace class, on ${\mathcal{H}} = {\mathcal{H}}_1 \otimes
{\mathcal{H}}_2$, such that \[ {\mathfrak{M}}(p) =
{\mathrm{Tr}}(Tp) \] for all finite rank projections $p$ in
${\cal P}({\cal H})$ if, and only if, ${\mathfrak{M}}$ is bounded on
one dimensional projection operators whose ranges are generated by
vectors in the algebraic tensor product ${\mathcal{H}}_1
\otimes_{{\mathrm{alg}}} {\mathcal{H}}_2$.


\begin{thebibliography}{99}
\bibitem{Wallach} {N.R.~Wallach}, \emph{An unentangled Gleason
theorem}, preprint, quant-ph/0002058.
\bibitem{Wright95} J.D.M.~Wright, \emph{The structure of decoherence
functionals for von
Neumann quantum histories,} J.~Math.~Phys.~\textbf{36} (1995), 5409-5413.
\bibitem{RudolphW97} {O.~Rudolph and J.D.M.~Wright},
\emph{On tracial operator representations of quantum decoherence
functionals},
J.~Math.~Phys.~{\bf 38} (1997), 5643-5652.
\bibitem{Wright98} J.D.M.~Wright, \emph{Decoherence functionals for
von Neumann quantum histories:
boundedness and countable additivity,} Comm.~Math.~Phys.~\textbf{191}
(1998), 493-500.
\bibitem{RudolphW98} {O.~Rudolph and J.D.M.~Wright},
\emph{The multi-form generalized Gleason theorem},
Comm.~Math.~Phys.~{\bf 198} (1998), 705-709.
\bibitem{RudolphW99} {O.~Rudolph and J.D.M.~Wright},
\emph{Homogeneous decoherence functionals in standard
and history quantum mechanics},
Comm.~Math.~Phys.~{\bf 204} (1999), 249-267.
\bibitem{Isham94} {C.J.~Isham},
\emph{Quantum temporal logic and decoherence functionals in the
histories approach to generalized quantum theory},
J.~Math.~Phys.~\textbf{35}
(1994), 2157-2185.
\bibitem{IshamLS94} {C.J.~Isham, N.~Linden and
S.~Schreckenberg},
\emph{The classification of decoherence
functionals: An analogue of Gleason's theorem},
J.~Math.~Phys.~\textbf{35} (1994), 6360-6370.
\bibitem{BunceW92a} L.J.~Bunce and J.D.M.~Wright,
\emph{The Mackey-Gleason problem},
Bull.~Amer.~Math.~Soc.~{\bf 26} (1992), 288-293.
\bibitem{BunceW92b} L.J.~Bunce and J.D.M.~Wright, \emph{Complex measures
on projections
in von Neumann algebras}, J.~London Math.~Soc.~\textbf{46} (1992), 269-279.
\bibitem{BunceW92c} L.J.~Bunce and J.D.M.~Wright,
\emph{The Mackey-Gleason problem for vector measures on projections
on von Neumann algebras},
J.~London Math.~Soc.~{\bf 49} (1994), 133-149.
\bibitem{Dvurecenskij93} A.~Dvure\v{c}enskij, \emph{Gleason's theorem
and its applications}, (Kluwer, Dordrecht, 1993).
\bibitem{Pedersen79} G.K.~Pedersen, \emph{$C^*$-algebras and their
automorphism groups}, (Academic, London, 1979).
\end{thebibliography}
\end{document}